\documentclass[12pt]{article}
\usepackage{graphicx}
\begin{document}
\begin{center}
{\Large On initial conditions for inflationary and bouncing cosmologies }\\
\bigskip
{\bf D.H. Coule}\\
\bigskip
Institute of Cosmology and Gravitation,\\ University of
Portsmouth, Dennis Sciama Building, Burnaby Rd., Portsmouth PO1 3FX, UK\\
\bigskip

\begin{abstract}

We consider the question of deriving initial conditions for scalar
fields that can drive  both an early and late, quintessence dark
energy, inflationary phase. Current notions of quantum cosmology
have difficulty in determining suitably displaced scalar fields
with vastly differing energy scales. Due to finite causal length
constraints the homogeneity of the dark energy field also presents
an unresolved {\em uniformity } problem.

 Some further specific concerns with  kinetic, phantom and assisted inflationary models
are outlined, especially when used as possible  dark energy
candidates.

 We review the use of the canonical measure in
predicting a single phase of inflation and find the negative
conclusions  of Gibbons and Turok can be allayed by means of a
{\em reasonable} quantum cosmological input.

 Further attempts at incorporating inflationary schemes into big bounce or cyclic
models are considered. Without some imposed low-entropy boundary
condition at the bounce any subsequent inflationary phase is
difficult to countenance despite claims to the contrary for loop quantum cosmology.


\end{abstract}
\end{center}
\newpage

{\bf 1.0 Scalar field driven cosmology}

Although inflationary cosmology is at present the established
model for the early universe  and is apparently in agreement with
observational evidence from WMAP [1] the underlying assumptions
are still far from being understood: essentially as we lack a full
quantum gravity description of the phenomena.

 There is also much speculation as to the possible cause of
 the apparent present acceleration of the universe. Various explanation
 have been postulated for this phenomena - see e.g.[2] for reviews.
 One possible model that has received much attention is to use a
 further, so-called quintessence, scalar field to drive an inflationary expansion, rather
 like that postulated for the early universe but now at a vastly
 reduced energy scale. One advantage of this scheme is that the
 two inflationary stages are caused by the same basic mechanism.
  We  later at times mention the alternative use of
 a  cosmological constant $\Lambda$ in driving the present
 acceleration - although problems with obtaining a suitable value
 for $\Lambda$ are well established [2,3].

 The present cosmological paradigm we then consider is outlined in
 Fig.(1),
  \vspace{1.0cm}\\
\hspace{4.0cm} ? $\rightarrow $ Inflation(1) $\rightarrow$
non-Inflation $
\rightarrow$ Inflation(2) $\rightarrow $ ?\\
\vspace{0.5cm}\\
Fig. 1: {\em Rough schematic of a model of the universe with, at
least, two inflationary phases. The preceding and subsequent
points of evolution are poorly understood. }\\
\vspace{1.0cm}

 where inflation(1) is caused by, say, a scalar field $\phi$ during
the early universe and inflation(2) a further field $\Phi$
dominating in the universe today.

 The simplest chaotic version of inflation uses a
displaced scalar field to violate the strong energy condition
[4-6]. During this time the potential $V(\phi)$ dominates over the
kinetic and spatial gradient terms. In the simplest FRW model the
energy density and pressure are given by e.g.[5,6]

\begin{equation}
\rho =\frac{\dot{\phi}^2}{2} +\frac {(\nabla \phi)^2}{2a^2} +
V(\phi)
\end{equation}
\begin{equation}
p=\frac{\dot{\phi}^2}{2} -\frac {(\nabla \phi)^2}{6a^2} - V(\phi)
\end{equation}

 If,
instead, initially the kinetic energy dominates it decays rapidly
as a stiff fluid $\dot{\phi}^2 \propto a^{-6} $, while a displaced
scalar field itself only decays logarithmically   slowly $\phi
\propto \ln(t)$: so one can expect an eventual inflationary phase
[4]. The spatial gradient $(\nabla\phi)^2$ term falls as $\sim
a^{-2}$ so again any initially displaced and slowly changing field
can be expected to eventually dominate. This spatial gradient
alone behaves roughly like a perfect fluid $p=(\gamma -1) \rho$
with $\gamma=2/3$ so is itself on the verge of inflationary
expansion, actually $a\sim t$ [7,8]. A single kinetic term behaves
like a perfect fluid with stiff equation of state $\gamma=2$. Note
that in scalar field driven cosmology one might argue that the
presence of a potential term $V(\phi)$ is actually necessary to
prevent an inhomogeneous scalar field simply producing an
everlasting coasting solution $a\sim t$.

The field $\phi$ has a different time dependence  when the
universe is dominated by a non-inflationary matter source, say
with radiation. For a massive scalar field case the solution of
the
\begin{equation}
\ddot{\phi}+3H\dot{\phi} =-m^2\phi
\end{equation}
field equation, with $H=1/2t$ for radiation, is given in the
slow-rolling limit by [9]
\begin{equation}
\phi\simeq\phi_i \exp(-\frac{1}{3}m^2t^2)
\end{equation}
The field says approximately constant for times $t<m^{-1}$. For
$m\sim 10^{-5}$, (the required value for perturbations [5,6])
inflation needs to proceed within the roll down time, now $10^{5}
t_{pl}$ in Planck units, if the field is still to be displaced
sufficiently from its minimum. This can impose a fine tuning of
the parameters if inflation(2) is also to be driven by a displaced
scalar field and is preceded by a long period of non-inflationary
behaviour.

 Arguments both from classical equipartition reasoning or quantum notions of likelihood
  have been advanced to
justify inflationary conditions  during the   early universe - see
e.g[5,6,10]. For example quantum cosmology might be able to
provide an initial large scalar field $\phi$ over an initial patch
of sufficient size $\sim H^{-1}$, with $H$ being the corresponding
Hubble parameter. This size is required since a negative pressure
is susceptible to any positive pressures surrounding it: a large
size allows more time for expansion to dominate before any
equalization processes can occur [7,11]. These arguments are far
from rigorous and few conclusions can be reached about the actual
likelihood of inflation happening without many underlying
assumptions.

These quantum arguments are often in contraction to the
requirements of a late universe inflationary stage. We wish to
point out this and a number of further difficulties
that then become apparent.\\

{\bf 2.0 Initial conditions for two subsequent $V(\phi)$
inflationary stages.}\\

In order for the field $\Phi$ to cause the present apparent
inflationary  acceleration it must be smooth over a patch size $L
>\sim H^{-1}_o$ where $H_o$ is the present Hubble parameter.
 To justify this {\em uniformity} one might like to make use of
the earlier inflation(1). This then requires that the field $\Phi$
be present during the earlier inflationary epoch and so displaced
from its minimum also beyond the initial patch size $H^{-1}$:
otherwise it would simply be red-shifted away just as a spatial
gradient. If instead the field $\Phi$ is produced later it cannot
be expected, without introducing a further horizon problem, to be
homogeneous over the present visible size of the universe, it
would have some smaller coherence length over which uniformity
could be justified - this would correspond to the particle horizon
size commencing from the time the field $\Phi$ was first produced.
There are related concerns if a cosmological constant $\Lambda$ is
postulated to drive inflation(2) but with the further issue that
symmetry breaking contributions during the early universe should
have vastly exceeded the required $\Lambda\sim 10^{-120}$ value
cf. section (2.6) in ref.[12].

 Topological defects might also be expected to form if the field
$\Phi$ is spontaneously broken  - see also sec. 28.3 in ref.[13]
for a related discussion. We ignore the case where inflation might
be formed within the cores of topological defects [14] - so
allowing instead $L<1/H$. This might be relevant for inflation(1)
but is unlikely to be suitable for inflation(2) - we know for
related ``universe in a lab'' work that the resulting inflationary
universe expands not within the existing space but into a new
evolving region [15]. Note that we generally  try and work within
a single Copernican principled model that obeys (near) spatial
homogeneity - so allowing the foliation as outlined in Fig.(1).

If the universe evolves from an initial Planck size nugget  a
quantum cosmological calculation using the Wheeler-DeWitt (WDW)
equation, with tunnelling boundary conditions, for a spatially
closed model can give an initial factor e.g. [5,6,10]

\begin{equation}
\Psi_T \sim \exp \left ( - \frac{1}{V(\phi) + V(\Phi) } \right )
\end{equation}
which is peaked at $V(\phi) +V(\Phi) \sim M^4_{pl}$. Such
arguments seem unlikely to explain the vast discrepancy in scale
between initial values for $V(\Phi)$ and $V(\phi)$. An alternative
Hartle-Hawking condition gives a corresponding $+$ sign in the
exponent and is seemingly less conducive to inflationary initial
conditions [10] - see however ref.[16] and section(3).

If the potential $V (\Phi)$ is taken to be a massive scalar field
i.e. $V(\phi)  =1/2M^2\Phi^2$ then the
 large Compton wavelength of the effective mass $M\sim 10^{-33} eV$ is
  of the order of the of the present size of the observable
  universe [2]. This might be inconsistent with the early
  universe's limited causal horizon size during inflation(1)
  cf.[17]. If the scalar field $\Phi$ is confined to regions of
  size $\sim m^{-1}$ there will be a large quantum uncertainty in
  the corresponding velocity $\dot{\Phi}$ which can dominate the
  dynamics and prevent the potential $V(\phi)$ causing inflation.

There are some alternative tracker potentials that considerably
reduce this initial discrepancy between $V(\phi)$ and $V(\Phi)$
but the requirement that the two inflationary stages should be
distinct still imposes the condition that initially
$V(\phi)>>V(\Phi)$. This discrepancy argument is therefore fairly
immune to the specific form of the quintessence  potential - see
e.g. [2,22-24] for various examples.

If we take $V(\phi)=m^2\phi^2$ and $m\sim 10^{-5}$ for
fluctuations, then to prevent $V(\Phi)$ inflating when
inflation(1) is finishing at $\phi\sim 1$ means that $V(\Phi)
<\sim 10^{-10} M^4_{pl}$. With the exponential function in eq.(5)
this is a large initial discrepancy to overcome. Roughly speaking
one is requiring Hartle-Hawking, or suppression,  like boundary
conditions for the $\Phi$ field and Tunnelling, or  enhancing,
ones for the high energy $\phi$ field.

Alternatively, one can consider quantum cosmology with compact
flat or open cosmologies [25-27], of compaction scale $L$,  then
the exponential suppression is lost and the action $S\sim
(aL)^3\sqrt{V(\phi)} $ [27] . Although we presently lack a
principle to impose a boundary condition in these cases, a small
action suggests that smaller $V(\phi)$ are preferred.  It still
remains unclear why two fields with vastly different actions can
be so produced.

There is also a possible complication that quantum fluctuations in
the field $\Phi$, having the same value as those in $\phi$ i.e.
$\delta \Phi \sim \delta \phi \sim H$, will cause diffusive
behaviour in the field $\Phi$. This can cause the field $\Phi$ to
grow to larger values so in turn allowing the potential $V(\Phi)$
to become dominant earlier than expected [28,29]. This has been
used as an argument to constrain the amount of inflation(1)
allowed, but by doing so it is in danger of counteracting  the
standard inflationary no-hair property: if too much or too little
inflation is a problem then it simply reintroduces  a further fine
tuning problem. A similar problem would occur if quantum
fluctuations are growing during the inflationary phase cf.[30].

The field $\Phi$ also has to be immune to being further jumbled up
during the turbulent reheating phase at the end of inflation(1):
so the two fields remaining essentially uncoupled.

{\bf 2.1 Extension to assisted inflation}

 A closely related form of inflation is
 assisted inflation where a number of fields $N$, each of which is
too steep to cause inflation by itself,  can increase the friction
so giving an overall inflationary expansion [31].  A quantum
cosmology argument would be required to see if such initially
displaced fields can be expected.  Because the potential of the
WDW equation is no longer necessarily  isolated from the origin at
zero scale factor the usual boundary conditions cannot give the
typical $\sim \exp(\pm 1/V(\phi)) $ factors -  somewhat similar to
the previously mentioned flat and open models. This also  occurs
with a classical signature change when the forbidden region is
also absent: the initial measure might then be uniform in $\phi$
[32]. We here ignore the N spatial gradient terms which is a
further complication.

 There seems a more serious problem, however,
if inflation(2) is to be driven by $N$ steep fields driven in
concert. During the previous non-inflationary phase the fields
still individually roll down the potential, so the corresponding
roll down time $t_r$ is still comparatively short for, say large
mass scalar fields where $t_r\propto m^{-1}$. For exponential type
potentials [33] that have no absolute minimum this would mean that
initially the fields are having to have very large potentials, say
at energy scale $\rho$ to provide some eventual inflationary
behaviour at a smaller energy scale $\rho_0$. But in this case
they should have provided an earlier inflationary phase back at
energy scale $\rho$, unless other matter fields present contrive
to prevent this. In summary, it appears less likely that assisted
inflation can provide inflation(2) due to this more severe fine
tuning.

{\bf 2.2 Phantom inflation}

 A further, and more extreme, type of inflation is caused  by
phantom matter where the equation of state gives $\gamma<0$ so
that the weak energy condition is also violated. This is a more
extreme pole-law expansion that produces a future big-rip
singularity: it therefore was generally discounted [34] for
inflation(1): one reason being that the fluctuations typically
have an unwanted  blue spectrum for increasing Hubble parameter
towards the impending singularity [34,35]. \footnote {More
recently in the limit of $\gamma<<0$ a near scale invariant
spectrum can also be obtained [36]; or else an additional scalar
might  source a slightly red perturbation spectrum [37]. Though
for $\gamma<-2/3$ there is possibly an insufficient no-hair
property due to a growing classical perturbation mode causing
inhomogeneity [38].} Also because the event horizon is contracting on approaching the big rip
the generalized second law suggests the collapsing solution is the actual physical solution-cf.section 4.0.

 A
simple example of phantom inflation  is to switch the sign of the
kinetic energy term [39]. However, the corresponding switch in the
spatial gradient term causes the spatial gradient to now
contribute a potentially dangerous positive pressure term cf.
eq.(2). In fact it contributes a term analogous to having a
positive curvature $k=+1$ present. The model is therefore more
susceptible to collapse than the standard scalar field model where
inhomogeneity still contributes a negative pressure e.g.[7]. One
might argue that if phantom is only being used for inflation(2),
this inhomogeneity can be suppressed by the earlier inflation(1),
but with the previous outlined provisos that we presently lack a
fully quantum version of the no-hair property.

A further difficulty is that phantom driven inflation requires the
presence of a potential that is driven up during the phantom phase
e.g.[40] . The starting value on the potential has to be near the
minimum so the phantom field climbs up the potential; this is  in
contrast to usual inflation where a large displaced potential is
required. Again if inflation(1) is scalar field driven and
inflation(2) phantom these complementary starting points on the
various potentials are problematic. If phantom is to be inflation(2)
 there is again a severe problem with fine tuning: since the
energy density of the phantom grows with scale factor the presence
of the earlier inflation should have expanded  the energy density
of the initial phantom, of say effective $\gamma=-1/3$,  by the
total increase in the scale
factor $\sim 10^{50} $ times.\footnote{ We assume that the scale factor grows a factor $\sim
10^{25}$ during each of  the inflationary and non-inflationary
phases.} The degree of fine tuning is therefore a factor $\sim
10^{50}$ times more than for the previous example using a massive
scalar field with standard inflation.

{\bf 2.3 Kinetic-inflation}

 A related example is k-inflation for either the early
 universe [41] or as dark energy [42]. One
includes a number of higher order derivative terms: so being in
some sense a generalization of the previous phantom case.   Some
kinetic terms still require negative signs in order to drive an
inflationary phase without the need of an explicit potential
$V(\phi)$ term.  However, instead  new arbitrary dimensional
constants have to be introduced to compensate dimensionally  for
the unusual higher order kinetic terms so obviating some of the
possible advantages of these models. Also, unlike in standard
potential driven inflation the corresponding spatial gradient
terms can potentially become the more dominant ones. Initially at
small initial scale factor the largest derivative term will
dominate. For a fourth derivative term the spatial derivative will
be $\pm (\nabla\phi)^4$ which contributes a term $\pm a^{-4}$.
Depending on the sign this is either a positive or negative
radiation term. A six derivative will give likewise a positive or
negative stiff fluid. Negative terms can push the model out of
bounds and restrict the generality of the corresponding  cosmic
no-hair property of such models [43].

One can also consider  2nd order derivative terms by means of an
arbitrary function of the D'Alembertian operator - so-called box
inflation [44]. This requires a closer analysis to see how spatial
gradient terms behave and whether it is compatible with
Ostrogradski's theorem - see e.g.[45].

With Born-Infeld type terms  an effective  square root on the
kinetic term is present see e.g.[46]. This causes  the
corresponding spatial gradient term to potentially only fall off
as $\sim a^{-1}$, so simulating a perfect fluid with inflationary
$\gamma=1/3$ equation of state. There is now a danger that a
suitably inhomogeneous field would instead cause perpetual spatial
gradient driven inflation. Note that in this limit the speed of
sound can diverge to infinity $c_s>>1$ [47], which might have
problems with causality [48].

Assuming an homogeneous field  we can consider some simplified
models of kinetic inflation. The kinetic Lagrangian or pressure
$p$ is given by a term [41],
\begin{equation}
p=F(X)
\end{equation}

with $X=\dot{\phi}^2/2$ for a spatially homogeneous field. For a
usual scalar field $p=X$. To take a particular example
\begin {equation}
p= aX+bX^2
\end{equation}
 In
order for the pressure to be somewhere negative one of the
constants $a$ or $b$ has to be taken negative. Since the energy
density is also of the form [41]
\begin{equation}
\rho=aX+3bX^2
\end{equation}
one can also get negative energy densities. In general the
quantities are related by an expression $\rho = 2Xp,_X-p$ with
comma representing derivative w.r.t $X$. When $p,_X=0$ there is a
possible de Sitter solution, $p=-\rho$. For the simple case
$a=-1=-b$ this occurs for $X=1/2$. The equation of state now
depends on the value of the kinetic energy. For large $X$ we get
in this case a radiation equation of state, while for $1/3<X<1/2$
there is phantom like behaviour. For $X<1/3$ the energy density is
negative. One can see this change by again looking at the scalar
field equation [41,46]
\begin{equation}
\ddot{\phi}+3Hc_s^2 \dot{\phi}=0
\end{equation}
where the speed of sound is defined as $c_s^2= (2X-1)/(6X-1)$.
Solving this equation the kinetic energy decays with scale factor
as  $\acute{\phi}^2 \propto
a^{-6c_s^2}$, showing how the kinetic energy behaves less stiff as $c_s^2$
is reduced and becoming de Sitter like as $c_s^2\rightarrow 0$.

This dependence on $X$ can be contrasted with usual potential
driven inflation where the equation of state depends on the slope
of the potential only: especially for an exponential potential.
The quantum boundary conditions try to impose   a large
displacement of the potential, and suppress the corresponding
kinetic energy [10]. Roughly speaking the kinetic inflation model
above having more variety is less suitable since, for example,
large initial $X$ might allow the universe to re-collapse before
the de Sitter value $X=1/2$ is ever approached. In the phantom
range, which might be unstable [49], the scalar field will be
driven up any scalar potential $V(\phi)$ present since in the slow
roll approximation the field equation takes the form, e.g.[46]
\begin{equation}
3p_{,X}H\dot{\phi} =-V'(\phi)
\end{equation}

A very cursory attempt at obtaining the corresponding WDW equation
for such a $F(X)$ finds that the resulting equation is a highly
non-linear wave equation with some aspects of the Boussinesq or
equation of transverse vibration e.g.[50]. There will be a number
of arbitrary constants to determine and as the solutions are so
dependent on the actual value of $X$ it will require a more
specific, than in the usual potential driven case, quantum
boundary condition proposal to make any real
predictions.\footnote{ An attempt [51] on obtaining the WDW
equation with a non-standard kinetic term has used classical
``on-shell'' approximations.} If kinetic inflation is to provide
inflation(2) the kinetic terms provide a extra, and probably
unwanted non-inflationary component in the early universe: or
generally with equation of state $p=\rho/(2n-1)$ for a $p=X^n$
term. Again it would be problematic for quantum cosmology to give
a displaced potential for inflation(1) together with suitable
kinetic terms to later drive inflation(2).

One can see the difficulty by considering the corresponding
Euclidean field equations for the expected instanton  when
inflationary behaviour is present. After an analytic continuation
of $t\rightarrow i\tau$ equation (9) effectively has the change
$X\rightarrow -X$, so again with solution $\acute{\phi}^2 \propto
a^{-6c_s^2}$ where now $c_s^2= (2X+1)/(6X+1)$ and prime denotes derivative w.r.t $\tau$. The relative sign
between $X$ and $X^2$ in equation(7) is lost and depending on
whether $X$ is large or small the model behaves simply  like that
of  a stiff fluid $p=\rho$ or radiation $p=\rho/3$ term. Although
this is suitable for the description of a quantum wormhole e.g[52],
typically $a\sim \cosh \tau$ and not a corresponding trig function
$a\sim cos(\tau)$  for the usual description of Euclidean
inflationary  behaviour.  Again this confirms that at the very
least rather ad hoc analytic continuations would be needed to
instead obtain a  DeSitter like instanton solution.

Back to the standard Lorentzian description: the kinetic driven
model, if used for inflation(1), anyway has to be amended since it
is attracted to the de Sitter value and would permanently inflate.
One tries to introduce functions of the field $\phi$ and allow
their evolution to change the energy density of the inflationary
phase. Because of gravitational wave constraints the energy
density must be below $\sim 10^{-10} M_{pl}^4$ around $\sim 40$
e-folding from the end of inflation [5,6]. A simplified factorized
version of this could be $p= K(\phi)F(X)$. For this case the
function $K(\phi)$ does not determine the actual equation of state
but $K(\phi)$ should be initially  displaced from its minimum so
further evolution can occur. This again will be difficult to
determine with quantum cosmological arguments since $K(\phi)$ is
not responsible for violating any energy conditions {\em per se}.
There is some sleight of hand with this model since it  still
depends on the values of $\phi$ even when an explicit potential
term is excluded: this can be quantified  using the canonical
measure where a further ambiguity due to $\phi$ will be
introduced.

Finally we should mention that the standard reheating mechanism which usually involves a
rapid oscillation around a potential's minimum needs to be replaced by some other reheating mechanism. Otherwise coupling of the kinetic energy to other fields would cause dissipation and the kinetic inflation rapidly end by
analogy to the usual reheating mechanism.

{\bf 3.0 Classical  measure for inflation}

One approach to determine the likelihood of an inflationary phase
is that of a classical canonical measure $\omega$ [53,54]. This
approach assumes the classical equations are valid up to arbitrary
energy scale $\rho$ and corresponds to a principle of indifference
being applied. For example, at  a fixed energy density the
corresponding value for $\omega$ is

 \begin{equation} \omega=-\dot{\phi}d\phi\wedge
d(a^3)
\end{equation}
 For such measures the probability of inflation is
found to be arbitrary, although the flatness problem could be
resolved for potentials unsuitable for inflation: so the flatness
problem did not strictly require inflation for its resolution.
However, in doing so the measure has to appeal to energy densities
vastly exceeding Planck values where the classical equations would
be expected to be superseded [55]. The early universe  is then
dominated by extremely large post-Planckian values of particularly
$\dot{a}$. This in turn sets the kinetic energy to be extremely
large in order to alone solve the flatness problem and give a
present energy density $\sim 10^{-30} gcm^{-3}$. See also [56,57]
for some further issues regarding the validity of this measure.

 Gibbons and Turok[58] wish to further resolve the ambiguity
 as to whether inflation occurs or not. Firstly, they have
 placed a cut-off for values of
the scale factor, or flatness $\Omega \sim 1$ that cannot be
distinguished experimentally. This seems to place  a rather
restrictive selection effect upon the measure. Unlike the simple
anthropic principle e.g.[59] observers are now having to decide
what they can or cannot measure.\footnote{ We ignore the presence
of inflation(2) in this section which should also ideally  be
incorporated as a requirement on the measure.}
 More subtle future experiments
might overcome this limitation. Indeed one might argue that in
order to resolve the flatness problem we should indeed consider
the universes arbitrary close to flatness and not simply remove
them  as being equivalent. Note also that although the canonical
measure can solve the flatness problem without inflation in
certain closed $k=1$  cases there is a further ambiguity for
bounded potentials: such as in the case of $R^2$ inflation [61,5]
when conformally transformed to an effective scalar field model
[55,61]. The measure $\omega$ diverges even for a fixed value of
the scale factor provided the initial energy density is
sufficiently large: above the plateau of the scalar potential.
Likewise for the factorized kinetic inflationary model with
$p=K(\phi)F(X)$, the measure at fixed $\dot{a}$, so signifying a
maximum closed universe before inflationary behaviour can proceed,
will involve a term $\omega\propto F(\phi) d\phi$, which without
restrictions on the form of $F(\phi)$  can contribute an infinity
of solutions due to the redundancy of the $\phi$ variable.

 More importantly for their argument[58] that the likelihood of inflation
 is largely suppressed they evolve backwards from
 the end of inflation and find the solution actually
 unstable to kinetic domination: or in general to the ``stiffest'' matter present.

  If we first accept this procedure there are a few ways to evade this
 conclusion. Firstly, in earlier loop quantum approaches the matter
  terms are affected by finite size corrections e.g.[62].
Massless scalar field can themselves violate the various energy
conditions and become, actually phantom-like,  inflationary. Then
the solution simply cannot evolve to any non-inflationary
behaviour in the past. This stage of loop driven inflation tends
to have insufficient duration without choosing arbitrary large
parameters and a second conventional phase of inflation was added
to the scheme [63]. This still suffers from insufficient likelihood of inflation [64] and using a phantom stage to prime a standard inflationary stage suffers from other issues of fine tuning [65].

 Another way of evading the scheme is in certain kinetic
inflationary models provided the previous ambiguity is first
resolved: when the generalized momentum $\pi=\dot{\phi} p,_{X}$
cannot diverge to infinity i.e. like the usual $\pi \propto
a^{-3}$ behaviour as the solution is evolved backwards without
pushing the corresponding energy density negative; or else the
universe evolves onto some previously collapsing phase cf.[66]. In
the previous model of section 2.3 this corresponds to taking $b$
negative - if on the contrary the momenta can diverge the
Gibbons-Turok argument still holds [67]. It might be argued that
both cases suggest inflation is unlikely but, with $b$ negative,
it also prevents the flatness problem from being solved since an
extremely large energy density is then not present to set the
initial value of $\Omega$ arbitrary close to unity.

 However it is well known that the
 inflationary solution is an attractor only in the forward
 direction, so  the field cannot be expected to evolve gradually up the
 potential as the solution is continued backwards. In the forward direction the inflationary
 solution is an attractor with the kinetic energy term decaying exponentially quicker than the
 value of the scalar field [68,69] - this is effectively  the no-hair property that exponentially suppresses
 any inhomogeneities present e.g.[7]. One can also see
 this difference in that particle horizons become event horizons
 and vice-versa when evolution is reversed [70]. So a backwards
 evolving inflationary solution has a corresponding particle
 horizon. This result can, though, be thought consistent with the requirement of
 inflation that the field $\phi$ be initially homogeneous over a
 length scale $L>1/H$. This is in some sense a highly ordered
 low-entropic state that requires some further justification. Evolving
 backwards one would expect to obtain, \`{a} la Gibbons and Turok,  a high entropy state that
 would indeed not be compatible with inflationary behaviour.

 This can give a highly likely probability of
 inflation provided the initial energy density, say $\rho_c$,  is taken sufficiently large
 [54,55]. If we represent the fraction of potential energy $V(\phi)$ to total energy by $F=V(\phi) /\rho_c$
 then, since $\dot{\phi} =\sqrt{\rho_c}(1-F)^{1/2}$, the measure becomes
 \begin{equation}
 \omega=M_{pl}^2(1-m^2\phi^2/M_{Pl}^4)^{1/2}d\phi \wedge d(a^3)
 \end{equation}
 where we use a massive scalar field of mass $m$ and take the
 energy density to have an initial Planck value.  We can integrate
 the expression over the variable $\phi$ to obtain for $m<<1$ a roughly uniform measure
 over $\phi$. By taking the ratios of field that allow sufficient
 numbers of e-folds of inflation to those that do not the
 probability of inflation is near unity only decreasing as
 $\rho_c$ is reduced.

  This can be thought of as a combination of classical and
 quantum reasoning since there is also an unbounded $\int d(a^3)$ integral
 over the scale factor to include: quantum here suggesting  small initial
 dimensions or action. Although in closed models the scale factor can be associated with a physical
 length in the case of open and flat models this is less clear how
 the formally infinite size can be associated  with the scale
 factor. This $a^3$ divergence was also responsible for solving
 the flatness problem without the need of inflation.

{\bf 3.1 Quantum measures for inflation}

As previously mentioned compared to the Tunneling boundary
condition
  the Hartle-Hawking wavefunction appears to suppress the initial potential
  energy as
$\Psi \sim \exp(1/V(\phi))$: although if one lets the field take
unrestricted values, way beyond Planckian, one can still get some
significant prediction for inflation [16]. Note however that the
Hartle-Hawking state also does not give a large value for the
kinetic energy that would have produced a singularity as
$a\rightarrow 0$. Although it seemingly agrees with Gibbons and
Turok in that inflation is exponentially suppressed it would not
alone be able to provide a solution alone to the flatness problem
without an explicit inflationary phase being present. We note that
the Hartle-Hawking boundary condition is somewhat ambiguous and
might also give big bang like solutions with exponential
potentials: having there both singular potential and kinetic
energies [71].

There is a further aspect: using a notion of  a typical boundary
condition Gibbons and Grischuk [72] found that Tunnelling like
boundary conditions were actually more typical: an indifference
principle was applied at initial Planckian values for the energy
density albeit with a simplified fixed  potential term. It remains
to be seen if this result can be uphold, especially with more
realistic inclusion of matter fields and  inhomogeneous modes.

The previous notion of a typical boundary condition was an example
of an {\em a priori} measure. The Hartle-Hawking wavefunction can
also be amended by assuming instead an observer's perspective
inside the present Hubble volume. One can firstly  insist that
$\phi>\phi_*$ for a Lorentzian space to develop [73,10]. Then
because of volume weighting, due to  a factor $\sim \exp(3N)$
caused by $N$ e-foldings of inflation, more Hubble volumes are
produced the longer inflation proceeds. This reasoning can
seemingly allow the Hartle-Hawking proposal to produce significant
inflation during our previous history [73,74]. But there is a
danger that this {\em a posteriori} reasoning could correct almost
any boundary condition proposal and make the distinction with
others e.g. the Tunnelling one irrelevant - it seemingly allows
quantum gravity processes to be influenced by future possible
events: teleological style reasoning. In any case after inflation
ceases the Hubble radius grows more rapidly than the scale factor
so eventually there is in any case only one Hubble patch: to
really produce more Hubble volumes one should consider flat and
open FRW models that formally have infinite size regardless of
inflation.

Given this argument  the universe is then found to bounce from a
previous collapsing phase. A closely related boundary condition,
roughly that the strong-energy condition be violated, has also
been formulated by Page [75], but this is only sufficient for
closed models: compact flat and open model would require  further
energy conditions to be violated.  The entropy $S\simeq 1/V(\phi)$
has to be low for inflation to ensue with large initial $V(\phi$
and so whether the collapsing phase can be allowed, perhaps with
its arrow of time reversed to prevent entropy build-up, or one
should consider the universe created at the bouncing point is
still unclear cf.[74]. There is also the suggestion, using a more
general complexified field, that the Hartle-Hawking proposal can
start the universe
 at its largest possible size, but perturbations
  apparently still grow during the subsequent collapse [76]. Note also that
previously a repulsive Planck potential[77]
had been introduced to  produce a bounce but this did not alone explain why a
subsequent inflationary stage was also present. We suspect the
more numerous ``separation like'' constants in kinetic
inflationary models could also allow bouncing like behaviour also
in non-closed models, but many unwanted and singular solution
would also have to be excluded.

An interesting  new development[78] is to start with a negative cosmological constant in Euclidean space and do an alternate
analytic continuation to obtain a Lorentzian universe now with a positive cosmological constant that could give inflationary expansion. Although the WDW equation has this symmetry for a FRW model with a simple scalar field we suspect more general matter sources, like radiation, will also potentially change sign. A preliminary investigation of Euclidean Schwartzschild-AdS with mass $M$ and charge $Q$ suggests the mass becomes imaginary $M\rightarrow iM$ and charge $Q^2\rightarrow
-Q^2$ with such a continuation to now Lorentzian Schwartzschild-deSitter space, so possibly limiting this approach if these general matter terms are initially present.

 {\bf 4.0 Bouncing or cyclic
universes}

 We can briefly consider further aspects of bouncing cosmologies where the
universe first collapses from a previous phase and in turn the
possibility of repeatedly using this mechanism to produce a cyclic
universe. The idealized model is outlined in Fig.(2).
\vspace{2cm}

 bounce $\rightarrow $ Inflation (1) $\rightarrow$ non-Inflation
$\rightarrow$ Inflation (2) $\rightarrow $ collapse \\
\vspace{0.5cm} $\uparrow$ \hspace{12.5cm} $\downarrow$

\hspace{4cm} $ \Longleftarrow $ Cyclic?
\\ \vspace {0.5cm}\\

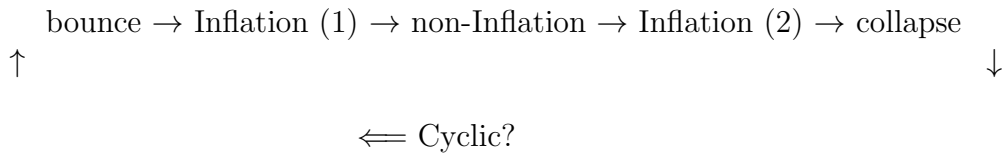
Fig.(2): {\em Possible extension of previous model to cyclic
behaviour by means of a suitable bounce. Can entropy be dissipated on going around the loop? }\\

 Consider the Friedmann equation for a FRW model [5,6]

\begin{equation}
H^2+\frac{k}{a^2} =\rho
\end{equation}

 A FRW bounce is
typically described by an equation of the form
\begin{equation}
H^2=\frac{A}{a^n} - \frac {B}{a^m}
\end{equation}

A bounce requires $m>n$ so the stiffer matter component requires
the minus sign. For a closed model the curvature plays this role
and only the strong energy has to be violated for a bounce to
happen - unlike the general case where more drastic violations are
required e.g.[79].

 Some approaches to quantum gravity suggest that
the Friedmann equation be modified such that
\begin{equation}
H^2=\rho-\frac{\rho^2}{\rho_c}
\end{equation}
where $\rho_c$ represents the critical energy scale.  This  occurs
 in more recent work in loop quantum gravity [80,81]. Related
behaviour might  be obtained with  brane models with an extra time
dimension [82] although this is probably observationally
discounted [83]; but see recently [84]. Note that a single
negative tension brane is not suitable: it differs from eq.(14) by
an overall minus sign on the R.H.S. since starting with a positive
5-dimensional Planck mass the negative tension causes the
4-dimensional Newton's constant to become negative cf.[85].

If one first  tries to work with non-inflationary matter and use
say a closed model  to re-collapse the universe one finds the
bounce size $a_b$ and maximum size $a_{max}$ do not differ
sufficiently. For the case of radiation $\ a_b^2 = a_{max}$ so it
is difficult to justify the universe becoming so large without
arbitrary large constants. To rectify this one would want to add
the inflation(1) phase  but again we have difficulties in
understanding how the strong energy condition becomes violated
after the bounce and not before.\footnote{ There are some bouncing
models e.g.[66] that permanently violate all the energy conditions
but these then have as much inflationary contraction as expansion:
so not contributing overall to resolving the various problems that
inflation is usually invoked for.} Indeed the previous results of
Gibbons and Turok now become relevant for a collapsing universe in
that the kinetic energy will increasingly dominate. We doubt
therefore that  that an ``anti-friction'' effect can drive  the
scalar field up the potential cf.[80] so that an inflationary
stage can proceed after the bounce. A more detailed account of this issue and  why the measure arguments[86], essentially applying eq.(12), should not be applied at the bounce in loop quantum cosmology and further entropic concerns are given in a separate paper [87]. We would just say here that although the apparent value of $F=V(\phi)/\rho_c$ for sufficient subsequent inflation appeared reasonable it corresponds to an entropy requirement $S\leq 10^{10}$ at the bounce and that the previous long classical evolution would  introduce dissipative behaviour so affecting any simple application of the canonical measure.

Because of these difficulties one might instead try to
resolve the various cosmological puzzles without an explicit
inflationary phase, or else impose some further  boundary
condition at the bounce itself like the amended Hartle-Hawking
one.

 Note also that for this modified Friedmann equation
$H\rightarrow 0$ as $\rho\rightarrow \rho_c$ so a large
cosmological constant is tending towards a static universe. This
incidentally can have some influence on whether quantum
fluctuations can produce eternal inflation cf.[5].

It has been noticed that this  Friedmann equation prevents a
phantom matter source $\rho\propto a^n$ with $n>1$ from reaching a
big rip singularity [88,89]. Instead the universe slows before
re-collapsing without the necessity of entering a high curvature
phase. With just a phantom matter source it will then approach a
super-collapsing phase. Previously there was a related model [90]
of the universe that started at the big rip before undergoing
super-collapse and eventually bouncing into a standard matter
dominated phase. The super-collapsing phase does not alone solve
the usual cosmological puzzles, for example the particle horizon
goes as:
\begin{equation}
R_H=a(t) \int _0^t \frac{dt}{a(t)} \propto t
\end{equation}
 for a collapsing scale factor $a\propto 1/t$, where $t=0$
 represents the start of the
 collapsing phase. This has the same behaviour as a usual
  non-inflationary expanding model. Neither does this
 collapsing phase reduce the entropy by fragmenting the universe
 which stems from a misuse of
 horizons and/or,  problematically to most people, equating the
 entropy with the corresponding universe's size cf.[89].

  More crucial is to obtain a generalized second law (GSL) e.g.[91]  of
 thermodynamics that allows entropy to increase together with a
 gap between the maximum allowed entropy and that actually present
 in the matter components [92]. Firstly, it is rather difficult to
 formulate a GSL , in an expanding model with phantom matter: one apparently has to
 introduce negative values for the entropy [93] or temperature
 [94]\footnote{ These large negative entropy/temperature values for
 the phantom are probably inconsistent with an
  inflation(1) phase}. Simply setting the entropy zero for the phantom
  component, like in an analogous  superfluid,
 would allow phantom matter to dissolve black holes upon approaching a
  big rip in violation of the GSL cf.[95].   Related negative entropy/temperature
   values have previously been suggested  for de Sitter [96,97],
 although the correct  sign of ``energy'' in the Gibbs equation
  confuses matters - see e.g.[98].

 Incidentally during the super-collapsing phase this problem of horizon entropy is
 obviated by the lack of an actual event horizon. But as we have
 previously discussed to obtain an ensuing inflation(1) phase
 requires a low-entropic state to develop. This is a rather
 difficult obstacle to overcome since the comoving entropy density
 would be expected to be growing, or at least remain constant, during the collapsing phase. It
 therefore appears difficult to obtain the cyclic universe as
 envisioned in the figure(2).

  Other approaches have tried to
 impose a cyclic structure but superimposed upon an underlying
 expanding universe. For example the quasi-steady state model [99] or
 the cyclic ekpyrotic one [100].
 These then attempt to use the cosmic no-hair
 property in order to dilute entropy production. However, this by
 sleight of hand
 introduces an infinity into which we can sweep the problem of excessive
 entropy production.\footnote{ There is a related suggestion of
 Penrose [101]
 that envisions an infinite conformal rescaling during the massless phase of a cosmological constant dominated universe
 to create suitable conditions of low-entropy for a subsequent big bang phase. However, the definition of
 allowed  entropy uses the notion of
 mass by means of the Planck units incorporated into Newton's constant $G$ - so this scheme at present
 appears somewhat inconsistent - unless gravity itself, and not simply mass, can be adjusted. }  It
 also means that all scales eventually originate from sub-Planck sizes of
 previous stages of the universe [102]  and further introduces
 geodesic-incompleteness problems of constantly expanding models
 [103].

{\bf 5.0 Conclusions}

 The presence of two inflationary stages  poses two sorts of
 problem: i) it shows up the weaknesses in  the original arguments
 that justified a displaced scalar, now  being in apparent contradiction with
  the necessary conditions of the
 second field; ii) a {\em uniformity} issue for the second scalar
 field is rather similar to that of the original horizon problem in
 non-inflationary models - so reintroducing a similar puzzle.

 The
general difficulty is that conditions for inflation(2) has also to
be set up before inflation(1) proceeds in order to have
homogeneous conditions over the present horizon size. Because of
finite particle horizon sizes it cannot simply be caused by
evolution from the end of inflation(1). There is a related problem
in obtaining sufficient  homogeneity for a   cosmological constant
if it is being used for inflation(2); but with the further
difficulty of  producing a sufficiently small value while various
phase transitions have taken place. Any early universe
inflation(1) can be ignored for the sake of adjusting  this
$\Lambda$ [12]: so the required uniformity is actually analogous
to a big bang model with an eventual $\Lambda$ dominated phase:
but such a model has a known horizon problem. Any  dark energy
inflationary stage is at the expense of an unnatural uniformity
which then requires a further explanation.

To put this another way, originally inflation was a single
assumption, hidden in the murky waters of Planck scale physics,
that solved a number of puzzles, but now the second phase of
inflation having  not yet ``solved'' anything is itself becoming a
puzzle requiring further explanation. This reasoning would become
worse the more often the universe stopped and started inflating,
i.e. if the ( inflation$\rightarrow$ non-inflation ) sequence in
Fig(1) was extended indefinitely in one or both directions. We
have tried to ignore more elaborate  notions of our universe
having branched off from an earlier, or still constantly evolving,
inflationary phase so that the actual universe would not have
distinct inflationary or non-inflationary stages as in the simple
model of Fig.(1). Firstly we do not think it realistic
observationally for inflation (2)to have formed this way. Secondly
other principles might constrain such branching phenomena, for
example in brane models the bulk space might fix the space-time
system to prevent quantum branching [104]; there might also be
trouble in obtaining a suitable arrow of time in any new quantum
dominated universe production [105].

 If the inflationary stages have mixed
causes, for example one being kinetic driven or phantom, similar
concerns are present. Generally speaking the various alternative
inflationary models: assisted, phantom, kinetic etc. appear less
suited to describe inflation(2), having even more fine tuning
concerns when a period of non-inflationary behaviour precedes
them.

Although we have considered only scalar field model these problems
should persist in many  higher derivative gravity models that have
been proposed as dark energy candidates [106,2]:  like those with
a Ricci scalar term $1/R^n$ added to the gravitational action
which can usually be transformed to a conformally equivalent
scalar field model. Some possible advantages of modifying gravity
schemes over the use of a quintessence field have been made in
ref.[45]; essentially the subsequent modified gravity inflationary
epoch is set {\em ab initio} into the action so obviating causal
constraints on obtaining a homogeneous quintessence field. The
problem is then displaced into the explanation of why the action
takes its particular form.  We would just add though, that such
higher derivative gravity theories especially with more general
Ricci tensor terms $R_{\mu\nu}R^{\mu \nu}$ or Weyl tensor are
known to have more limited cosmic no-hair properties - with
possible premature collapse [107] or anisotropic inflationary
solutions [108]. This might not be a serious problem for
inflation(2) since we do not necessarily want to establish the
cosmological principle into the far distant future but it is
unattractive if the inflation is of this limit form compared
perhaps with inflation(1); or it requires starting conditions that
only slightly depart from FRW in order to restrict the effects of
these more general higher derivative terms involving combinations
of Ricci, Weyl or  Riemann tensors. Such modified gravity theories
also tend to be strongly constrained by unwanted consequences
during the early universe - see e.g.[109].

To summarize  some possible avenues for future study:\\
$\bullet$ A suitable boundary condition that can give a { \em a
priori} prediction of two distinct stages of inflation, either
starting from some creation event or from a previous pre-big bang
phase. Presently the usual proposals are too energy density
dependent, not amenable to justifying initial conditions at vastly
differing energy scales.
\\
$\bullet $  Quantum formulism of cosmological no-hair property to
explain possible smoothing of dark energy scalar field, or else
the initial non-causal like  {\em uniformity}
issue needs to be further resolved.  \\
$\bullet $ Quantization with higher derivative scalar matter terms
i.e. kinetic or Box inflation: obtaining solutions of WDW equation
together
with a justifiable  boundary condition that eliminates unwanted solutions.\\
 $\bullet$ Can a entropy sink be incorporated to
produce an actual cyclic model. Various ideas of e.g. infinite
spatial size, reversing arrow of time in collapsing model, to
dilute entropy do not appear realistic. Use of algorithmic complexity/entropy to help clarify different
models[110].  Perpetually expanding
models typically have a  incompleteness problem: although various
counterexamples appear possible [111,112]. Neither do such cyclic
models explain why the entropy at any time is not already
maximized by the presence of black holes. \\
$\bullet$ Can modified gravity models provide adequate
inflationary stages: both in the early and late universes? Can the
specific action be justified from more fundamental principles and
what restrictions on initial conditions are still necessary for
its implementation?

 {\bf Acknowledgement}\\ I should like to thank A. Aguirre, O. Corrandi,
W.  Nelson, D. Page, Yun-Song Piao,  A. Vikman  and Yi Wang for
helpful discussions.
\newpage

{\bf References}\\
\begin{enumerate}
\item D.N. Spergel et al, Astrophys. J. Suppl. 170 (2007) p.377.\\
arXiv:0603449.[astro-ph]\\
http://map.gsfc.nasa.gov
\item P.J.E. Peebles and B. Ratra, Rev. Mod. Phys. 75 (2003)
p.559.\\
R.R. Caldwell, Phys. World 17 (2004) p.37.\\
S.M. Carroll, preprint astro-ph/0310342.\\
T. Padmanabhan, preprint gr-qc/0503107.\\
E.J. Copeland, M. Sami and  S. Tsujikawa,  Int. J. Mod.  Phys. D
15 (2006)
p.1753.\\
 P. Ruiz-Lapuente, Class.
Quant. Grav. 24 (2007) p.R91.
\item S. Weinberg, Rev. Mod. Phys. 61 (1989) p.1.
\item A.D. Linde, Phys. Lett. B 129 (1983) p. 177. {\em ibid}
175 (1986) p.395.
\item A.D. Linde, ``Particle Physics and Inflationary cosmology''
(Harwood Press)  1990.
\item E.W. Kolb and M.S. Turner, ``The Early Universe''
(Addison-Wesley: New York) 1990.
\item D.S. Goldwirth and T. Piran, Phys. Rep. 214 (1992) p.223.
\item M.S. Madsen and P. Coles, Nucl. Phys. B 298 (1988) p. 2757.
\item M.S. Madsen, Class. Quant. Grav. 7 (1990) p.2073.
\item J.J. Halliwell, in ``Quantum cosmology and baby universes''
eds. S. Coleman et al. (World Scientific: Singapore) 1991.\\ D.L.
Wiltshire, in  ``Cosmology the physics of the universe'', eds. B.
Robson et al. (World Scientific, Singapore) 1996. \\ also as
preprint gr-qc/0101003.
\item T. Vachaspati and M. Trodden, Phys Rev. D 61 (2000)
p.023502.
\item R. Bousso, Gen. Rel. Grav. 40 (2008) p.607.
\item R. Penrose, ``The Road to Reality'', Johnathon Cape: London
(2004).

\item A.D.  Linde, Phys. Lett. B 327 (1994) p.208.\\
A. Vilenkin, Phys. Rev. Lett. 72 (1994) p.3137.
\item S. Blau, E. Guendelman and A.H. Guth, Phys. Rev. D 35 (1987)
p.1747.\\
E. Fahri and A.H. Guth, Phys. Lett. B 183 (1987) p.149.\\
E. Fahri, A.H. Guth and J. Guven, Nucl. Phys. B 339 (1990)
p.417.\\
W. Fischler, D. Morgon and J. Polchinski, Phys. Rev. D 41 (1990)
p.2638.\\
A.D. Linde, Nucl. Phys. B 372 (1992) p.421.\\
S. Ansoldi and E. I. Guendelman, Prog. Theor. Phys. 120 (2008) p.985.
\item D.N. Page, Phys. Rev. D 56 (1997) p.2065.
\item P.C.W. Davies, Int. J. Theor. Phys. 47 (2008) p.1949.      \\
Da-Ping Du, Bin Wang and Ru-Keng Su, Phys. Rev. D 70 (2004)
p.064024.
\item G.W. Gibbons and S.W. Hawking, Phys. Rev. D 15 (1977)
p.2738.
\item R.M. Wald, Phys. Rev. D 28 (1983) p.2118.
\item A.K. Raychaudhuri and B. Modak, Class. Quant. Grav. 5 (1989) p.225.\\
{\em see also}:
 P. Anninos, R.A. Matzner, T. Rothman and M.P. Ryan, Phys.
Rev. D 43 (1991) p. 3821.
\item A.A. Coley, S. Hervik and W.C. Lim, Phys. Lett. B 638 (2006)
p.310.
\item P.J. Steinhardt, L. Wang and I. Zlatev, Phys. Rev. D 59
(1999) p.123504.
\item T. Barreiro, E.J. Copeland and N.J. Nunes, Phys. Rev. D 61
(2000) p.127301.
\item R.R. Caldwell and E.V. Linder, Phys. Rev. Lett. 95 (2005)
p.141301.

\item Y.B. Zeldovich and A.A. Starobinsky, Sov. Astron. Lett. 10
(1984) p.135.
\item D.H. Coule and J. Martin, Phys. Rev. D 61 (2000) p.063501.
\item A.D. Linde, JCAP 0401 (2004) p.004.
\item M. Malquarti and A.R. Liddle, Phys. Rev. D 66 (2002) p.023524.
\item J. Martin and M. Musso, Phys. Rev. D 71 (2005) p.063514.
\item Chun-Hsien Wu, Kin-Wang Ng and L.H. Ford, Phys. Rev. D 75 (2007) p.103502.
\item A.R. Liddle, A. Mazumdar and F.E. Schunck, Phys. Rev. D 58
(1998) p.061301.
\item J. Martin, Phys. Rev. D 49 (1994) p.5086.
\item F. Luchin and S. Matarrese, Phys. Rev. D 32 (1985) p.1316.\\
J.J. Halliwell, Phys. Lett. B 185 (1987) p.341.\\
A.B. Burd and J.D. Barrow, Nucl. Phys. B 308 (1988) p.929.

\item D.H. Coule, Phys. Lett. B 450 (1999) p. 48.
\item S. Mollerach, S. Matarrese and F. Lucchin, Phys. Rev. D 50
(1994) p.4835.
\item Yun-Song Piao and E. Zhou, Phys. Rev. D 68 (2003)
p.083515.\\
M. Baldi, F. Finelli and S. Matarrese, Phys. Rev. D 72 (2005)
p.083504.
\item Yun-Song Piao and Yuan-Zhong Zhang, Phys. Rev. D 70 (2004)
p.063513.
\item J.C. Fabris and S.V.B. Goncalves, Phys. Rev. D 74 (2006)
p.027301.\\
K.A. Bronnikov, J.C. Fabris and S.V.B. Goncalves, J. Phys. A 40 (2007) p.6835.
\item R.R. Caldwell, Phys. Lett. B 545 (2002) p.23.
\item V. Faraoni, Class. Quant. Grav. 22 (2005) p.3235.
\item C. Armendariz-Picon, T. Damour and V. Mukhanov, Phys. Lett.
B 458 (1999) p.209.
\item  C. Armendariz-Picon, T. Damour and V. Mukhanov, Phys. Rev.
D 63 (2001) p.103510.
\item A.D. Rendall, Class. Quant. Grav. 23 (2006) p.1557.
\item A. Anisimov, E. Babichev and A. Vikman, JCAP 0506 (2005)
p.006.
\item R.P. Woodard, Lect. Notes Phys. 720 (2007) p.403.
\item Wei Fang, H.Q. Lu and Z.G. Huang, Class. Quant. Grav. 24 (2007) p.3799.
\item V. Mukhanov and A. Vikman, JCAP 0602 (2006) p.004.
\item C. Bonvin, C. Caprini and R. Durrer, Phys. Rev. Lett. 97
(2006) p.081303.\\
G.F.R. Ellis, R. Maartens and M. MacCallum, Gen. Rel. Grav. 39
(2007) p.1651.
\item L. Raul Abramo and N. Pinto-Neto, Phys. Rev. D 73 (2006)
p.063522.
\item A.D. Polyanin and V.F. Zaitsev,  Handbook of nonlinear partial differential
equations, Chapman and Hall, Boca Raton (2004).
\item H.Q. Lu, Z.G. Huang, W. Fang and P.Y. Ji, Eur. Phys. J. C55 (2008) p.329.
\item A. Carlini, D.H. Coule and D.M. Solomons, Mod. Phys. Lett. A11 (1996) p.1453.
\item G.W. Gibbons, S.W. Hawking and J.M. Stewart, Nucl. Phys. B
281 (1987) p.736.\\
{\em see also:} V.A. Belinski and I.M. Khalatnikov, Sov. Phys. 66
(1988) p.3.
\item S.W. Hawking and D.N. Page, Nucl. Phys. B 298 (1988) p.789.
\item D.H. Coule, Class. Quant. Grav. 12 (1995) p.455.
\item S. Hollands and R.M. Wald, Gen. Rel. Grav. 34 (2002)
p.2043.\\
{\em ibid} hep-th/0210001.
\item L. Kofman, A.D. Linde and V. Mukhanov, JHEP 0210 (2002)
p.057.
\item G.W. Gibbons and N. Turok, Phys. Rev. D 77 (2008) p.063516.
\item J.D. Barrow and F.J. Tipler, The Anthropic Cosmological
Principle, Oxford University Press: Oxford 1986.
\item A.A. Starobinsky, Phys. Lett. B 91 (1980) p.99.
\item D.N. Page, Phys. Rev. D 36 (1987) p.1607.
\item M. Bojowald, Living Rev. Rel. 8 (2005) p.11
\item M. Bojowald, Phys. Rev. Lett. 89 (2002) p.261301.
\item C. Germani, W. Nelson and M. Sakellariadou, Phys. Rev. D 76 (2007) p.043529.
\item D.H. Coule, Class. Quant. Grav. 22 (2005) p.R125.
\item L.R. Abramo and P. Peter, JCAP 0709 (2007) p.001.
\item Miao Li and Yi Wang, JCAP 0706 (2007) p.012.
\item D.S. Salopek and J.R. Bond, Phys. Rev. D 42 (1990) p.3936.\\
\item D.S. Goldwirth, Phys. Lett. B 256 (1991) p.354.
\item W. Rindler, ``Essential Relativity 2nd edn.'' (
Springer-Verlag: New York) 1977.
\item D.N. Page, in ``The Future of Theoretical Physics and
Cosmology'' eds. G.W. Gibbons, E.P.S. Shellard and S.J. Rankin,
Cambridge University Press 2003. \\{\em also } hep-th/0610121.

\item G.W. Gibbons and L.P. Grishchuk, Nucl. Phys. B 313 (1989)
p.736.
\item J.B. Hartle, S.W Hawking and T. Hertog, Phys. Rev. Lett. 100 (2008) p.201301.
\item S.W. Hawking, arXiv:0710.2029.
\item D.N. Page, Class. Quant. Grav. 25 (2008) p.154011.
\item D. Green and W.G. Unruh, gr-qc/0206068.
\item H.D. Conradi and H.D. Zeh, Phys. Lett. A 154 (1991) p.321.\\
H.D. Conradi, Phys. Rev. D 46 (1992) p.612.
\item J.B. Hartle, S.W. Hawking and T. Hertog, arXiv: 1205.3807.\\
J.B. Hartle, S.W. Hawking and T. Hertog, arXiv: 1207.6653.
\item C. Molina Paris and M. Visser, Phys. Lett. B 455 (1999)
p.90.

\item P. Singh, K. Vandersloot and G.V. Vereshchagin, Phys. Rev. D
74 (2006) p. 043510.
\item A. Ashtekar, gr-qc/0702030.
\item Y. Shtanov and V. Sahni, Phys. Lett. B 557 (2003) p.1.\\
M.G. Brown, K. Freese and W.H. Kinney, astro-ph/0405353.
\item G. Dvali, G. Gabadadze and G. Senjanovic, hep-ph/9910207.
\item I. Quiros,  arXiv:0706.2400
\item C. Barcelo and M. Visser, Phys. Lett. B 482 (2000) p.183.
\item A. Ashtekar and D. Sloan, Phys. Lett. B 694 (2010) p.108.\\
A. Corichi and A. Karami, Phys. Rev. D 83 (2011) p.104006.\\
A. Ashtekar and D. Sloan, Gen. Rel. Grav. 43 (2011) p.3619.
\item D.H. Coule, arXiv:0802.1867.
\item M. Sami, P. Singh and S. Tsujikawa, Phys. Rev. D 74 (2006)
p.043514.
\item L. Baum and P.H. Frampton, Phys. Rev. Lett. 98 (2007)
p.071301.\\
P.H. Frampton, astro-ph/0612243.
\item F.G. Alvarenga and J.C. Fabris, Class. Quant. Grav. 12
(1995) p.L69.\\
{\em see also:}  A.B. Batista, J.C. Fabris and S.V.B. Goncalves,
Class. Quant. Grav. 18 (2001) p.1389.
\item P.C.W. Davies and T.M. Davis, Found. Phys. 32 (2002) p.1877.
\item J.D. Barrow, New. Astron. 4 (1999) p.333.
\item G. Izquierdo and D. Pavon, Phys. Lett. B 633 (2006) p.420.\\
 J.A.S. Lima and  J.S. Alcaniz, Phys. Lett. B 600 (2004) p.191.\\
 I. Brevik, S. Nojiri, S.D. Odintsov and L. Vanzo, Phys. Rev. D
70 (2004) p.043520.
\item P.F. Gonzalez-Diaz and C.L. Sigueza, Phys. Lett. B 589
(2004) p.78.
\item G. Izquierdo and D. Pavon, Phys. Lett. B 639 (2006) p.1
\item M.D. Pollock and T.P. Singh, Class. Quant. Grav. 6 (1989)
p.901.
\item D. Klemm and L. Vanzo, JCAP 0411 (2004) p.006.
\item T. Padmanabhan, Phys. Reports 406 (2005) p.49.\\
M. Spradlin, A. Strominger and A. Volovich, hep-th/0110007.
\item F. Hoyle, G. Burbidge and J.V. Narlikar, ``A different
approach to cosmology'' ( Cambridge University Press: Cambridge)
1999.

\item P.J. Steinhardt and N. Turok, Phys. Rev. D 65 (2002)
p.126003.
\item R. Penrose, Before the big bang? lecture at The Newton Institute Cambridge 2005.\\
 available online: www.newton.ac.uk
\item D.H. Coule, Int. J. Mod. Phys. D 12 (2003) p.963.
\item A. Borde, A.H. Guth and A. Vilenkin, Phys. Rev. Lett. 90
(2003) p.151301.
\item D.H. Coule, Gen. Rel. Grav 36 (2004) p.2095.
\item  B. McInnes, Phys. Rev. D 77 (2008) p.123530.\\
arXiv:0711.1656.
\item S.M. Carroll, V. Duvvuri, M. Trodden and M.S. Turner, Phys.
Rev. D 70 (2004) p.043528.
\item A.L. Berkin, Phys. Rev. D 44 (1991) p.1020.
\item J.D. Barrow and S. Hervik, Phys. Rev. D 74 (2006) p.
124017.\\
{\em ibid}, D 73 (2006) p.023007.\\
D. Muller and S.D.P. Vitenti, Phys. Rev. D 74 (2006) p.083516.

\item A.A. Starobinsky, JEPT Lett. 86 (2007) p.157.
\\ {\em and references therein}
\item P.C.W. Davies, Int. J. of Theor. Phys. 28 (1989) p.1051\\
W.H. Zurek, Phys. Rev. A 40 (1989) p.4731.
\item G. Ellis and R. Maartens, Class. Quant. Grav. 21 (2004)
p.223.
\item A. Aguirre and S. Gratton, Phys. Rev. D 65 (2002) p.083507.\\
A. Aguirre, arXiv:0712.0571.[hep-th]

\end{enumerate}
\end{document}